# Robust Glioblastoma Segmentation and Volumetry Without T2-FLAIR: External Validation of Targeted Dropout Training

Marco Öchsner, MB BChir[1]; Lena Kaiser, PhD[2]; Robert Stahl, MD, PhD[1]; Nathalie L. Albert, MD[2,3,4]; Thomas Liebig, MD[1]; Robert Forbrig, MD[1]; Jonas Reis, MD[1]

[1] Institute of Neuroradiology, LMU University Hospital, LMU Munich, Germany

[2] Department of Nuclear Medicine, LMU University Hospital, LMU Munich, Germany

[3] German Cancer Consortium (DKTK), Partner site Munich, Munich, Germany

[4] Bavarian Cancer Research Center (BZKF), Munich, Germany

**Corresponding Author**

Jonas Reis, MD

Institute of Neuroradiology

LMU Hospital, Ludwig Maximilian University

Marchioninistraße 15

81377 Munich (Germany)

Tel. +49 89 4400 72626

Fax +49 89 4400 72509

Email: jonas.reis@med.lmu.de

## Abstract

**Objectives:** To externally validate targeted T2 fluid-attenuated inversion recovery (T2-FLAIR) dropout for robust automated glioblastoma segmentation and whole-tumor volumetry without T2-FLAIR, while preserving performance when the full MRI protocol is available.

**Materials and Methods:** In this retrospective multi-dataset study, 3D nnU-Net models were developed on BraTS 2021 (n=848) and externally validated on an independent University of Pennsylvania glioblastoma cohort (n=403). Models were trained with or without targeted T2-FLAIR dropout zeroing the T2-FLAIR channel during training. Testing used prespecified T2-FLAIR-present and T2-FLAIR-absent scenarios; the absent scenario was simulated by zeroing the T2-FLAIR channel at inference. The primary endpoint was per-patient overall region-wise Dice similarity coefficient (DSC). Secondary endpoints were region-specific DSC, 95th percentile Hausdorff distance and Bland-Altman whole-tumor volume bias.

**Results:** In external validation, performance was preserved with the full MRI protocol: overall median DSC was 94.8% (interquartile range [IQR] 90.0%-97.1%) with dropout and 95.0% (IQR 90.3%-97.1%) without dropout. In the T2-FLAIR absent scenario, targeted dropout improved overall median DSC from 81.0% (IQR 75.1%-86.4%) to 93.4% (IQR 89.1%-96.2%). Whole-tumor DSC improved from 60.4% to 92.6%, whole tumor 95th percentile Hausdorff distance from 17.24 mm to 2.45 mm, and whole-tumor volume bias from -45.6 mL to 0.83 mL.

**Conclusions:** In an independent external test cohort, targeted T2-FLAIR dropout preserved glioblastoma segmentation performance with the full MRI protocol and substantially reduced whole-tumor segmentation error and volumetric bias when T2-FLAIR was absent. These findings support targeted sequence dropout as a practical robustness strategy for automated glioblastoma analysis in retrospective and heterogeneous clinical workflows.

**Key Words:**

Glioblastoma; magnetic resonance imaging; fluid-attenuated inversion recovery; tumor segmentation; volumetry; external validation; missing imaging sequences; robustness.

## Introduction

Glioblastoma is the most common primary malignant brain tumor in adults and requires MRI assessment of contrast-enhancing and non-enhancing components for treatment planning and response assessment (1,2). Because manual delineation of tumor compartments is time-consuming and subject to inter-reader variability, reliable automated tumor segmentation is clinically attractive (3-5). Deep learning methods, particularly U-Net-style architectures and self-configuring models such as nnU-Net, have achieved strong performance and are increasingly used as the standard tool for brain tumor segmentation (6,7). The Multimodal Brain Tumor Segmentation (BraTS) challenges have accelerated this development through large, multi-institutional multiparametric MRI datasets with standardized annotations (8,9). However, most high-performing pipelines still require a complete four-sequence structural MRI protocol (T1-weighted, post-contrast T1-weighted (T1-CE), T2-weighted, and T2-fluid-attenuated inversion recovery (T2-FLAIR) (4,6) despite the fact that retrospective and real-world clinical data are often incomplete, heterogeneous, or partially unusable.

This deployment gap matters because the likelihood of missing data increases with the number of required inputs, and prior work suggests that whole-tumor segmentation depends strongly on T2/FLAIR information (10-12). T2-FLAIR therefore merits specific study rather than a generic missing-modality discussion. Although RANO 2.0 assigns less weight to non-enhancing disease in many *IDH*-wildtype glioblastoma response-assessment settings, T2/FLAIR remains relevant for characterizing non-enhancing abnormality and edema and for imaging contexts in which non-enhancing disease is clinically consequential, including antiangiogenic treatment (2,13). From a segmentation perspective, absent or non-diagnostic T2-FLAIR may therefore distort whole-tumor delineation and volumetry even when the other structural sequences are available. Several strategies have been proposed to address missing MRI modalities, including synthesis of absent sequences using generative models (11,14,15), hetero-modal representation learning (16), and sequence dropout training (17-18). Among these, targeted sequence dropout is appealing for clinical deployment because it does not require an additional synthesis model or separate models for each possible input subset. The intended use of the present approach is robust segmentation-driven tumor delineation and whole-tumor volumetry when T2-FLAIR is missing or unusable in retrospective or heterogeneous clinical MRI, rather than replacement of diagnostic image interpretation or a universal solution to all missing-modality scenarios. What remains insufficiently defined is whether targeted T2-FLAIR dropout can preserve segmentation performance under the full four-sequence protocol while preventing clinically meaningful failure when T2-FLAIR is unavailable on independent external testing. We therefore developed models on BraTS 2021 and evaluated all primary study endpoints in the withheld UPenn-GBM cohort.

## Materials and Methods

### Study Design and Reporting

This retrospective multi-dataset study analyzed only publicly available, de-identified MRI datasets. Accordingly, ethics approval and informed consent were not required, and the study involved no interaction with human participants. Reporting was aligned with the Checklist for Artificial Intelligence in Medical Imaging (CLAIM) and its 2024 update, with a completed checklist provided in the Supplement (19,20).

### Cohort and Reference Standard

Model development used the RSNA-ASNR-MICCAI BraTS 2021 training cohort after excluding the University of Pennsylvania glioblastoma cohort (UPenn-GBM, remaining BraTS n=848) (9,21). Within BraTS 2021, the released images had been preprocessed with skull stripping, co-registration, and 1-mm isotropic resolution, and expert annotations of tumor subregions were provided. Independent testing used the UPenn-GBM cohort from BraTS 2021 and was withheld entirely from model development (n=403), also comprising baseline pre-treatment MRI and expert-revised tumor segmentations (21,22); the workflow is illustrated in Figure 1. We included all cases with (i) an available structural MRI including at least T1, T1-CE, T2, and T2-FLAIR, and (ii) an available reference standard segmentation. In both cohorts, the reference standard segmentation comprised necrotic/non-enhancing tumor (NCR/NET), peritumoral edema (ED), and enhancing tumor (ET) provided by the dataset creators. No study-specific manual contour editing was performed (9,21).

**Segmentation Targets and Inference Scenarios**

Two output configurations were evaluated: Region-wise outputs were prespecified for the main analyses because they correspond more directly to clinically interpretable tumor compartments: whole tumor (WT = NCR/NET ∪ ED ∪ ET), tumor core (TC = NCR/NET ∪ ET), and enhancing tumor (ET). Label-wise outputs following the original BraTS convention (NCR/NET, ED, ET) were analyzed as supportive secondary outputs. Each external test case was evaluated under two prespecified inference scenarios: T2-FLAIR-present (T1, T1-CE, T2 and T2-FLAIR) and T2-FLAIR-absent. The unavailable-sequence condition was simulated by replacing the preprocessed T2-FLAIR channel with zeros immediately before inference. This paired design isolates the effect of absent T2-FLAIR while holding anatomy, reference segmentation, and the remaining input sequences constant. It was chosen to evaluate a specific deployment setting, robust segmentation-driven tumor delineation and whole-tumor volumetry when T2-FLAIR is missing or unusable in retrospective or heterogeneous clinical MRI, rather than replacement of diagnostic image interpretation or a universal solution to arbitrary missing-modality patterns. No synthetic image generation was used, and no attempt was made to model degraded or artifact-contaminated T2-FLAIR.

**Model Development**

All models were developed with the 3D full-resolution nnU-Net framework (7). Six model configurations were trained: {label-wise, region-wise} × {T2-FLAIR dropout rate (r)=0.0, 0.35, 0.50}. For the T2-FLAIR dropout models, only the T2-FLAIR channel was replaced with zeros during training, with probability r per training sample, while T1, T1-CE, and T2 were never dropped. Training used patient-level 5-fold cross-validation (80% training / 20% validation within each fold) on the BraTS 2021 development cohort with nnU-Net default 3D full-resolution settings and 1,000 epochs per fold. For each fold, the checkpoint with the highest mean foreground Dice similarity coefficient on the validation partition was selected, and cross-validation ensemble learning was used for inference. Fold-level validation was used solely for model optimization and ensemble construction and was not treated as a separate study endpoint. No UPenn case was used for training, hyperparameter tuning, or model selection.

**Statistical Analysis**

The primary endpoint was per-patient overall region-wise DSC on external testing. Secondary endpoints were region-specific DSCs (WT, TC, ET), supportive label-wise DSCs (NCR/NET, ED, ET), 95th percentile Hausdorff distance (HD95) and whole-tumor volumetric agreement. Volumetric agreement was evaluated with Bland-Altman analysis as predicted minus reference whole-tumor volume in milliliters. For paired model comparisons, effect sizes are reported as median paired differences with 95% bootstrap confidence intervals from 2,000 patient-level resamples. The equivalence of T2-FLAIR-present performance was assessed using two one-sided tests (TOST) with prespecified bounds of ±1.5 DSC percentage points. The equivalence margin was chosen a priori to represent a clinically negligible difference relative to the larger performance shifts expected in the simulated T2-FLAIR-unavailable scenario (23). All analyses were performed in Python 3.12 using PyTorch 2.8.0, NumPy 2.1.3, SciPy 1.15.3, and pandas 2.2.3.

## Results

Results are reported as median [Q1-Q3] across patients. All primary results are derived from independent external validation on the UPenn-GBM cohort. Main external-validation results for the region-wise models are summarized in Table 1 and 2 and Figures 2-4. Figure 4 quantifies the corresponding cohort-level effect on whole-tumor volumetric agreement, and Figures 3 and 4 show representative segmentation examples. Supportive label-wise analyses, additional paired equivalence testing and sensitivity analyses are provided in the Supplement (Supplementary Tables S1-S5).

**Performance with T2-FLAIR present**

In the independent external cohort, when T2-FLAIR was available, targeted T2-FLAIR dropout training preserved segmentation performance relative to standard four-sequence training (Table 1). For the region-wise models, overall DSC was 95.0% [90.3-97.1] without dropout and 94.8% [90.0-97.1] with targeted dropout (r=0.35), with comparable boundary error (HD95 1.61 mm [1.14-2.72] vs 1.55 mm [1.14-2.74]). Region-specific performance was similarly stable, including WT DSC of 95.5% versus 95.6%, ET DSC of 94.1% versus 94.0%, and TC DSC of 96.5% versus 96.4% for r=0.0 and r=0.35, respectively. Equivalence of r=0.35 versus r=0.0 within ±1.5 DSC percentage points was supported across endpoints in the T2-FLAIR-present setting (TOST $p<0.001$ throughout; Table 2).

**Performance and whole tumor volumetric agreement in the simulated T2-FLAIR-unavailable Scenario**

In the same external cohort, the simulated T2-FLAIR-unavailable scenario, performance deteriorated substantially without robustness training, whereas targeted dropout preserved high performance (Table 1). Overall DSC improved from 81.0% [75.1-86.4] without dropout to 93.4% [89.1-96.2] with targeted dropout (r=0.35). The main gain was in WT segmentation: WT DSC improved from 60.4% [45.7-71.9] to 92.6% [88.7-95.2], and WT HD95 improved from 17.24 mm [11.66-24.20 mm] to 2.45 mm [1.73-5.00 mm] (Table 1). Bland-Altman analysis showed marked systematic whole-tumor underestimation without dropout, with mean volume bias (predicted minus reference standard) of -45.6 mL (95% bootstrap confidence interval [CI], -48.8 to -42.6 mL) and wide limits of agreement from -108.0 to 16.8 mL (Figure 2A). The r=0.35 dropout model largely eliminated this bias, with a mean volume bias of 0.83 mL (95% CI, -0.28 to 1.92 mL) and tighter limits of agreement from -20.8 to 22.5 mL (Figure 2B). Representative qualitative examples of this failure mode and its recovery are shown in Figures 3 and 4. By contrast, ET and TC segmentation were comparatively stable, with ET DSC of 92.4% versus 93.5% and TC DSC of 94.4% versus 96.2% without versus with targeted dropout. Supportive label-wise analyses localized this failure mode to edema, with marked recovery under targeted dropout (Supplementary Table S1). Equivalence was not supported for r=0.35 versus r=0.0 in the T2-FLAIR-absent setting (Table 2), reflecting differences far exceeding the ±1.5 percentage-point equivalence margin.

**Supplementary analyses**

Results for the alternative dropout configuration (r=0.50), additional paired equivalence testing, and contextual benchmarking against HD-GLIO are provided in the Supplementary Tables and were consistent with the main findings.

## Discussion

In this retrospective multi-dataset study with external validation, targeted T2-FLAIR dropout preserved glioblastoma segmentation performance under the complete four-sequence MRI protocol and substantially improved robustness when T2-FLAIR was unavailable. The clearest translational benefit was in whole tumor delineation and whole-tumor volumetry, with supportive label-wise analyses localizing this effect primarily to edema. That pattern is important because no single segmentation metric adequately reflects clinical usability: concordant improvement across Dice similarity coefficient, 95th-percentile Hausdorff distance, and Bland-Altman volume bias supports the interpretation that the observed gain was clinically meaningful rather than merely numerical. The appropriate claim is therefore narrow: targeted T2-FLAIR dropout is a practical robustness measure for a prespecified missing-sequence scenario, with likely relevance to unusable T2-FLAIR acquisitions, but not a universal solution for incomplete or degraded MRI (7,9,19,21).

These findings align with and extend prior work on missing-modality brain tumor segmentation. Existing approaches include modality synthesis, hetero-modal representation learning, and sequence dropout training (11,14,16,24,25). Among these, targeted dropout has practical advantages because it does not require a separate synthesis network or multiple subset-specific models (18). Our results also accord with evidence that FLAIR-containing input combinations carry much of the information needed for whole-tumor segmentation, whereas contrast-enhanced T1-weighted imaging contributes mostly to enhancing-tumor delineation (18). More broadly, prior work has shown that automated glioblastoma segmentation can be made more robust to heterogeneous routine clinical MRI and to missing sequences, supporting the view that incomplete-protocol imaging is a genuine deployment problem rather than a purely technical edge case.

Recent work further underscores this translational priority. In pediatric brain tumors, dropout-based robustness training and synthesis-based strategies have recently been evaluated under missing-MRI conditions, highlighting the same core problem in a different disease setting (26). In parallel, contrast- and resolution-agnostic frameworks such as TumorSynth have been introduced for brain tumor segmentation across heterogeneous clinical MRI inputs (27). Our findings complement this emerging literature by focusing on a narrower but common adult glioblastoma failure mode: absent T2-FLAIR. In an independent external cohort, a simple targeted dropout strategy preserved performance when full MRI protocol was available and substantially reduced whole-tumor undersegmentation and volumetric bias when T2-FLAIR was unavailable.

The observed error pattern was neuroradiologically plausible and clinically interpretable. Edema and non-enhancing tumor components are predominantly characterized as T2/T2-FLAIR hyperintensity, so a model trained exclusively on complete four-sequence inputs can implicitly learn a reliance on T2-FLAIR for whole-tumor delineation and underuse complementary cues from T2-weighted imaging when T2-FLAIR is absent. Targeted T2-FLAIR dropout repeatedly exposes the network to a mixture of complete and T2-FLAIR-absent inputs during training, encouraging reliance on redundant information in the remaining contrasts and reducing sensitivity to a single missing channel. The relative stability of enhancing tumor segmentation across conditions is consistent with its stronger dependence on post-contrast T1-weighted imaging (2,9,10,13).

In practical terms, the value of the present approach lies in preserving quantitative whole-tumor assessment when protocol completeness cannot be assumed. T2/FLAIR abnormality remains relevant for pretreatment tumor characterization, delineation of edema and non-enhancing abnormality, radiotherapy planning, radiomics, and whole-tumor volumetry (13). This broader relevance is also consistent with recent amino-acid PET literature: PET RANO 1.0 formalized PET-based response assessment in diffuse gliomas, and in newly diagnosed glioblastoma, [$^{18}$F]FET PET frequently identifies metabolically active tumor burden that extends beyond MRI contrast enhancement and may remain measurable when MRI-RANO does not show measurable disease (28,29). In the present study, a standard four-sequence-trained model showed marked whole-tumor underestimation when T2-FLAIR was absent, whereas targeted dropout largely eliminated this bias. The most immediate application is therefore robust segmentation-driven volumetry in retrospective datasets, multicenter studies, and heterogeneous clinical imaging workflows, rather than replacement of diagnostic image interpretation.

This study has several limitations. First, the T2-FLAIR-unavailable condition was simulated by zeroing an otherwise available preprocessed channel. This paired design isolates the effect of absent T2-FLAIR but does not capture degraded-yet-present acquisitions, including motion corruption, partial coverage, misregistration, or unusual intensity characteristics. Second, although the external test cohort was excluded from model development, it remained a single-institution (multi-scanner) cohort drawn from the broader BraTS/UPenn public dataset, which tempers claims of generalizability. Third, the analysis was restricted to a targeted missingness pattern and to baseline pretreatment glioblastoma; robustness to other missing sequences, combinations of missing and degraded inputs, postoperative examinations, and follow-up imaging remains unproven. Fourth, the reference standard was derived from public expert annotations, and no study-specific contour adjudication or inter-reader variability analysis was performed. Fifth, we did not directly compare the proposed strategy against a dedicated three-sequence model trained without T2-FLAIR, so the present results demonstrate the practical value of a single robust model rather than superiority over modality-specific deployment strategies. Finally, the HD-GLIO comparison is best interpreted as contextual benchmarking rather than a definitive head-to-head comparison because training data, preprocessing, label definitions, and intended use differ across studies.

Consequently, future work should evaluate the approach under more realistic missingness and degradation patterns, across additional external cohorts with genuinely missing or non-diagnostic T2-FLAIR, and against dedicated three-sequence training. Such comparisons would also help clarify the practical trade-off between a single robustness-trained model and modality-specific deployment strategies.

In conclusion, in independent external validation on the UPenn-GBM cohort, targeted T2-FLAIR dropout training within a 3D nnU-Net model preserved glioblastoma segmentation performance when the full MRI protocol was available and substantially improved robustness in a simulated T2-FLAIR-unavailable scenario. The main benefit was reduction of clinically relevant whole-tumor failure, including undersegmentation of edema and systematic whole-tumor volumetric bias. These findings support targeted sequence dropout as a practical robustness strategy for automated glioblastoma MRI analysis in retrospective datasets and heterogeneous clinical imaging workflows in which T2-FLAIR may be missing or unusable.

## Declaration of generative AI and AI-assisted technologies in the manuscript preparation process

During the preparation of this work the authors used ChatGPT (OpenAI; GPT-5.2) for language editing in order to improve clarity and scientific tone. After using this tool/service, the authors reviewed and edited the content as needed and take full responsibility for the content of the published article.

## Data and Code sharing statement

No new imaging data were generated for this study. The MRI data and reference standard segmentations used for model development and external testing are available from public repositories under their respective data use agreements, including BraTS 2021 and UPenn-GBM. The authors did not redistribute these datasets. All code used for preprocessing, training, inference, and statistical analysis is stored in a GitHub repository and will be made available upon acceptance. GitHub repository: https://github.com/LMU-NRAD/FLAIR-dropout. Release status: private during peer review, to be made public upon acceptance. License: Apache-2.0

## Declaration of competing interest

The authors declare no conflicts of interest.

## Funding Statement

This study did not receive any funding.

## Ethical Statement

This retrospective multi-dataset study analyzed only publicly available, de-identified MRI datasets. Accordingly, ethics approval and informed consent were not required, and the study involved no interaction with human participants.

## Tables

| | Overall | | Whole Tumor | | Enhancing Tumor | | Tumor Core | |
|---|---|---|---|---|---|---|---|---|
| T2-FLAIR drop for training | DSC med [Q1-Q3] | HD95 med [Q1-Q3] | DSC med [Q1-Q3] | HD95 med [Q1-Q3] | DSC med [Q1-Q3] | HD95 med [Q1-Q3] | DSC med [Q1-Q3] | HD95 med [Q1-Q3] |
| *Segmentation with FLAIR input* | | | | | | | | |
| r=0.0 | 95.0 [90.3-97.1] | 1.61 [1.14-2.72] | 95.5 [91.9-97.8] | 1.73 [1.00-3.32] | 94.1 [87.6-96.8] | 1.00 [1.00-1.41] | 96.5 [92.9-98.1] | 1.41 [1.00-2.45] |
| r=0.35 | 94.8 [90.0-97.1] | 1.55 [1.14-2.74] | 95.6 [91.9-97.8] | 1.62 [1.00-3.32] | 94.0 [87.5-96.8] | 1.00 [1.00-1.41] | 96.4 [92.9-98.0] | 1.41 [1.00-2.45] |
| r=0.5 | 94.8 [90.2-97.0] | 1.61 [1.14-2.71] | 95.3 [91.9-97.7] | 1.73 [1.00-3.46] | 93.7 [87.4-96.7] | 1.00 [1.00-1.41] | 96.5 [92.8-98.1] | 1.41 [1.00-2.45] |
| *Segmentation without FLAIR input* | | | | | | | | |
| r=0.0 | 81.0 [75.1-86.4] | 7.59 [5.31-11.21] | 60.4 [45.7-71.9] | 17.24 [11.66-24.20] | 92.4 [86.3-95.6] | 1.00 [1.00-1.73] | 94.4 [89.8-96.7] | 2.45 [1.73-4.58] |
| r=0.35 | 93.4 [89.1-96.2] | 2.02 [1.41-3.41] | 92.6 [88.7-95.2] | 2.45 [1.73-5.00] | 93.5 [87.5-96.4] | 1.00 [1.00-1.41] | 96.2 [92.4-97.8] | 1.73 [1.00-2.83] |
| r=0.5 | 93.7 [89.1-96.2] | 2.03 [1.41-3.46] | 93.0 [89.4-95.7] | 2.24 [1.41-4.64] | 93.7 [87.4-96.5] | 1.00 [1.00-1.41] | 96.2 [92.5-97.8] | 1.73 [1.00-2.83] |

**Table 1. UPenn-GBM external validation (n=403): Region-wise segmentation performance with T2-FLAIR present and absent at inference.** Performance of region-wise nnU-Net models (WT, TC, ET) trained on BraTS 2021 with targeted T2-FLAIR dropout probability rates of 0.0, 0.35, and 0.50 and evaluated under two prespecified inference scenarios: T2-FLAIR present (all four sequences provided) and T2-FLAIR absent (T2-FLAIR channel replaced with zeros). Values are median [Q1-Q3] across patients. Overall denotes the unweighted macro-average across WT, TC, and ET.

**Abbreviations:** DSC = Dice similarity coefficient (reported as %); HD95 = 95th-percentile Hausdorff distance (mm); WT = whole tumor; TC = tumor core; ET = enhancing tumor; FLAIR = fluid-attenuated inversion recovery; r = dropout probability rate.

| | Median DSC r(0.35)-r(0.0) [95%CI] | TOST p-Value |
|---|---|---|
| *Segmentation with T2-FLAIR input* | | |
| Overall | -0.017 [-0.046, 0.014] | <0.001 |
| ET | -0.04 [-0.094, -0.006] | <0.001 |
| TC | -0.011 [-0.040, 0.015] | <0.001 |
| WT | -0.005 [-0.029, 0.023] | <0.001 |
| *Segmentation without T2-FLAIR input* | | |
| Overall | 11.35 [10.63, 12.21] | **1** |
| ET | 0.75 [0.663, 0.837] | **0.98** |
| TC | 1.12 [0.97, 1.29] | **1** |
| WT | 31.1 [28.2, 33.48] | **1** |

**Table 2. UPenn-GBM external validation (n=403): Paired equivalence testing (TOST) for r=0.35 versus r=0.0 in region-wise models.** Region-wise models trained with targeted T2-FLAIR dropout (r=0.35) were compared with no-dropout models (r=0.0) under T2-FLAIR-present and T2-FLAIR-absent inference scenarios. Entries report median paired DSC differences (r=0.35 − r=0.0) in percentage points with 95% bootstrap confidence intervals. TOST p values test equivalence within prespecified bounds of ±1.5 DSC percentage points (α=0.05).

**Abbreviations:** TOST = two one-sided tests; DSC = Dice similarity coefficient; WT = whole tumor; TC = tumor core; ET = enhancing tumor; FLAIR = fluid-attenuated inversion recovery; r = dropout probability rate.

## Figures

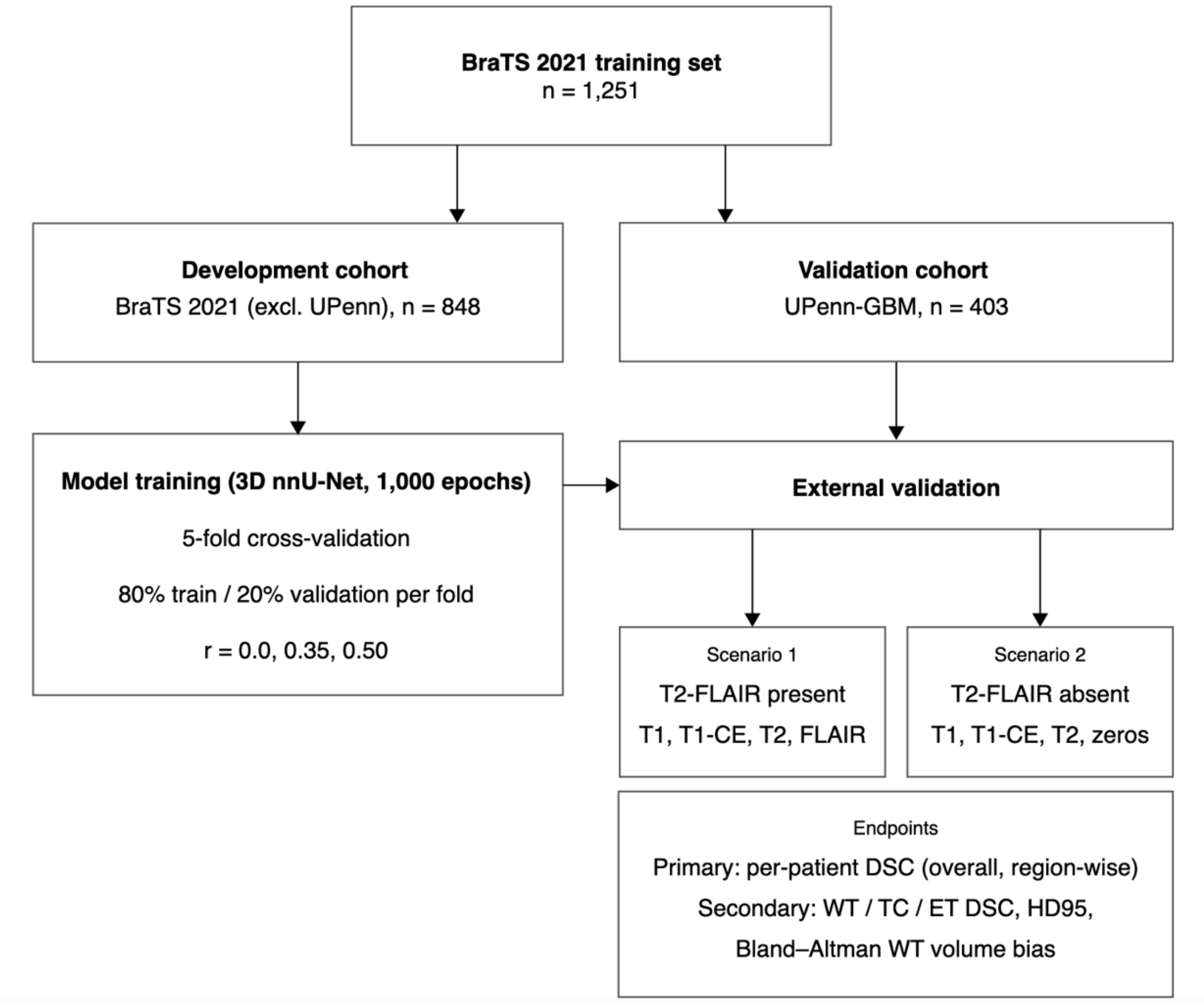

**Figure 1. Study design and evaluation workflow.** BraTS 2021 cases excluding UPenn were used for five-fold cross-validation training of 3D nnU-Net models with targeted T2-FLAIR dropout probability rates of 0.0, 0.35, and 0.50; external validation used the held-out UPenn-GBM cohort under two prespecified inference scenarios: T2-FLAIR present and T2-FLAIR absent (T2-FLAIR channel replaced with zeros). The primary endpoint was per-patient Dice similarity coefficient (DSC); secondary endpoints were region-wise whole-tumor (WT), tumor core (TC), and enhancing tumor (ET) DSC, 95th-percentile Hausdorff distance (HD95), and Bland-Altman WT volume bias.

**Abbreviations:** DSC = Dice similarity coefficient; ET = enhancing tumor; FLAIR = fluid-attenuated inversion recovery; HD95 = 95th-percentile Hausdorff distance; TC = tumor core; WT = whole tumor; r = dropout rate.

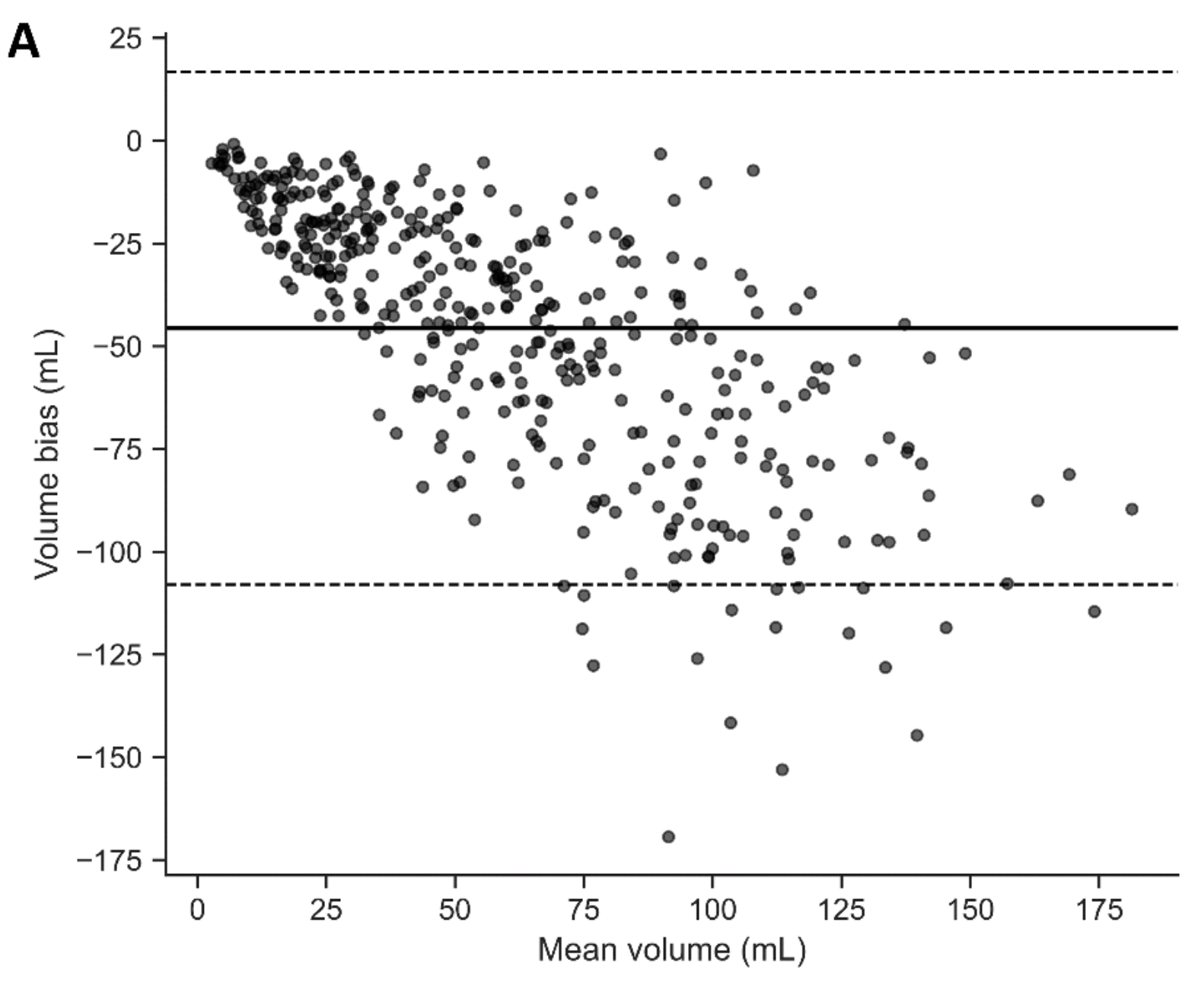

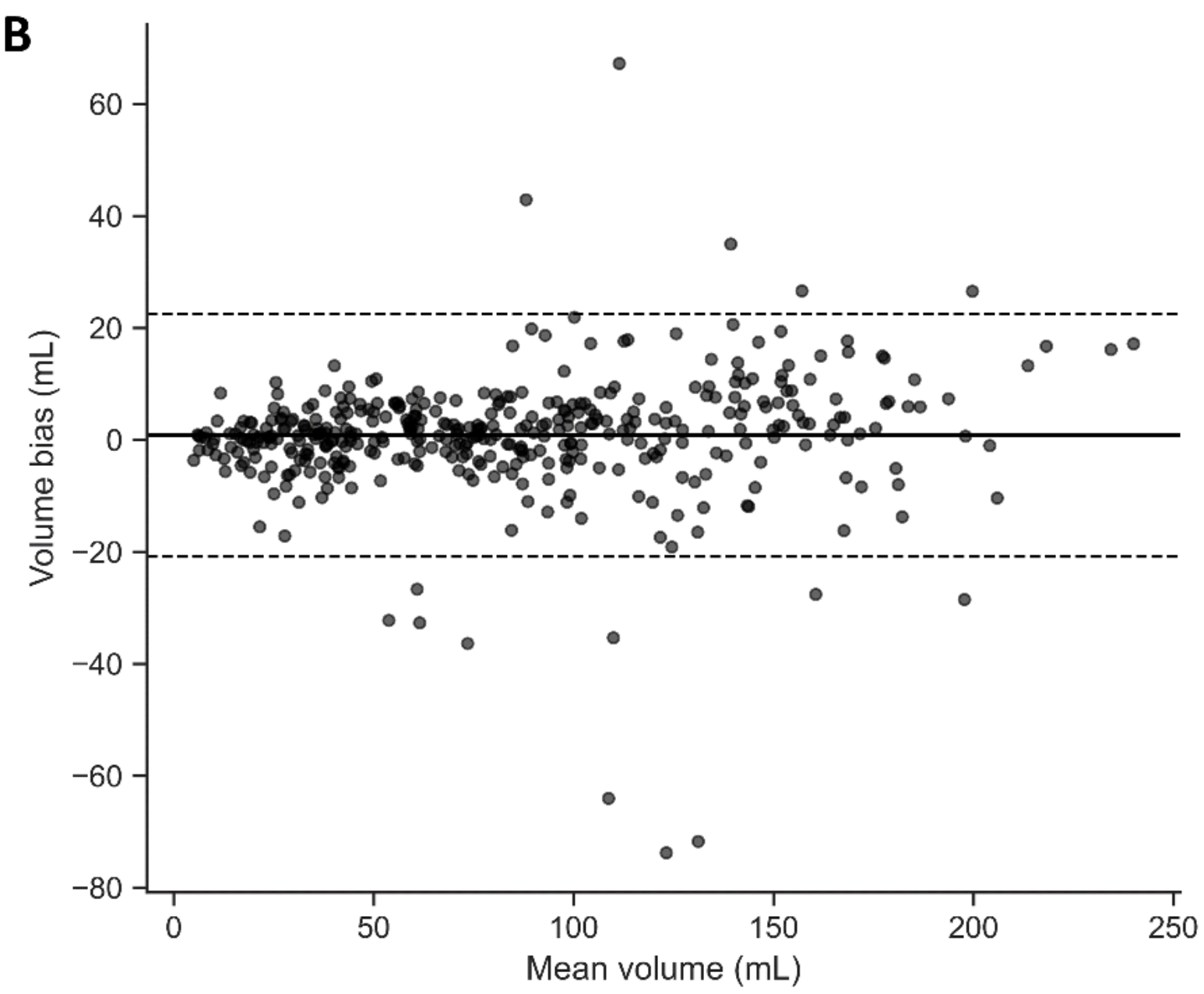

**Figure 2. External validation on UPenn-GBM (n=403): Restoration of whole-tumor volumetric agreement without T2-FLAIR.** Bland-Altman plots of predicted minus reference standard whole-tumor volume (mL) in the T2-FLAIR-absent scenario for the region-wise nnU-Net trained **(A)** without dropout (r=0.0; Bias: -45.59 mL; Limits of Agreement: -107.96, 16.78 mL) and **(B)** with targeted T2-FLAIR dropout (r=0.35; Bias: 0.83 mL, Limits of Agreement: -20.83, 22.49 mL). Solid lines indicate mean bias; dashed lines indicate 95% limits of agreement.

**Abbreviations:** FLAIR = fluid-attenuated inversion recovery; r = dropout rate.

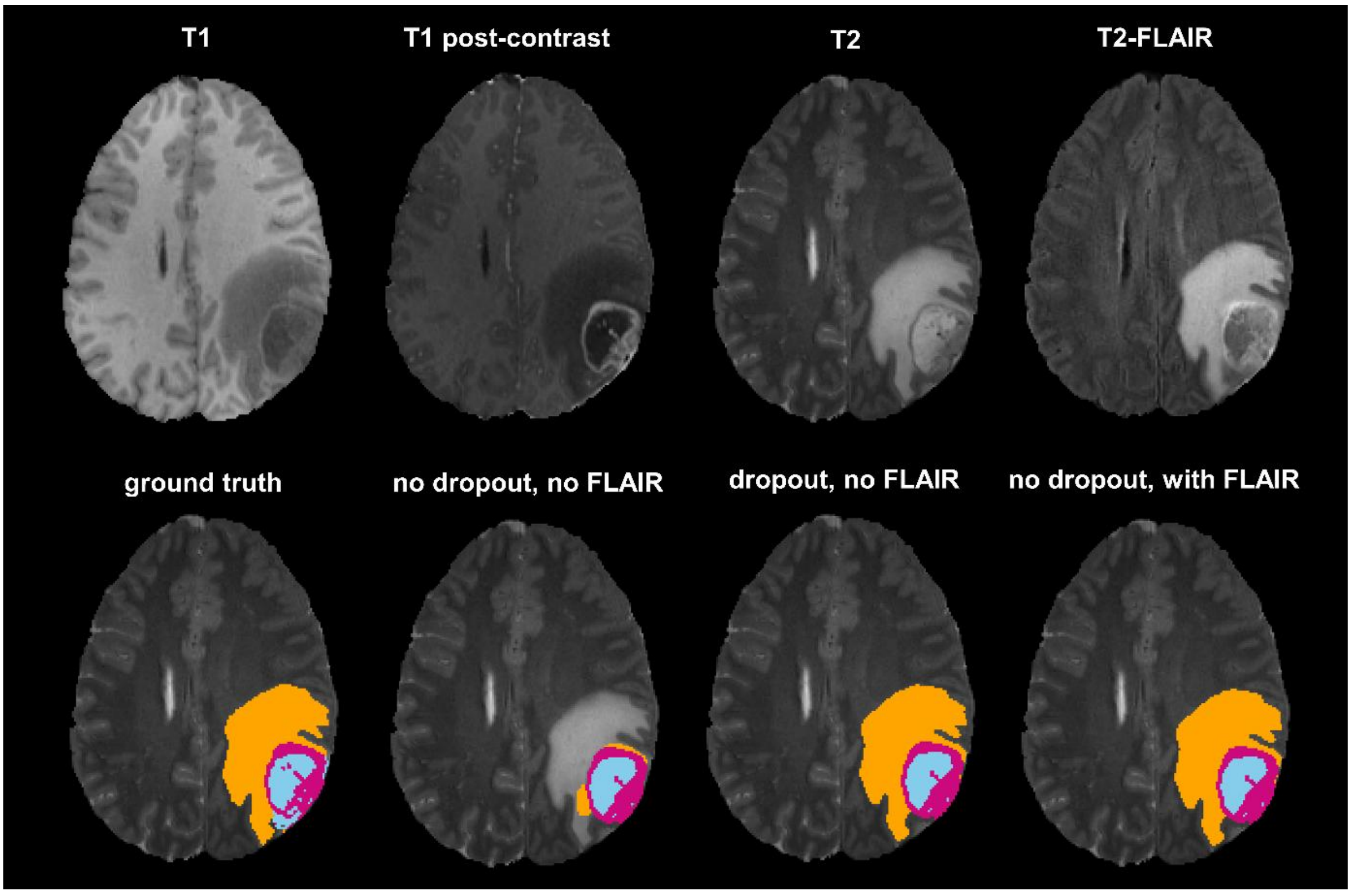

**Figure 3. Single representative external validation case of whole-tumor rescue without T2-FLAIR.** Top row shows T1-weighted, post-contrast T1-weighted, T2-weighted, and T2-FLAIR images. Bottom row shows the reference standard, the region-wise nnU-Net prediction without dropout (r=0.0) in the T2-FLAIR-absent scenario, the prediction with targeted T2-FLAIR dropout (r=0.35) in the same scenario, and the no-dropout prediction in the T2-FLAIR-present scenario. This case illustrates whole-tumor undersegmentation without dropout when T2-FLAIR is absent and recovery with targeted dropout, corresponding to the cohort-level volumetric effect summarized in Figure 2.

**Abbreviations:** FLAIR = fluid-attenuated inversion recovery; r = dropout rate.

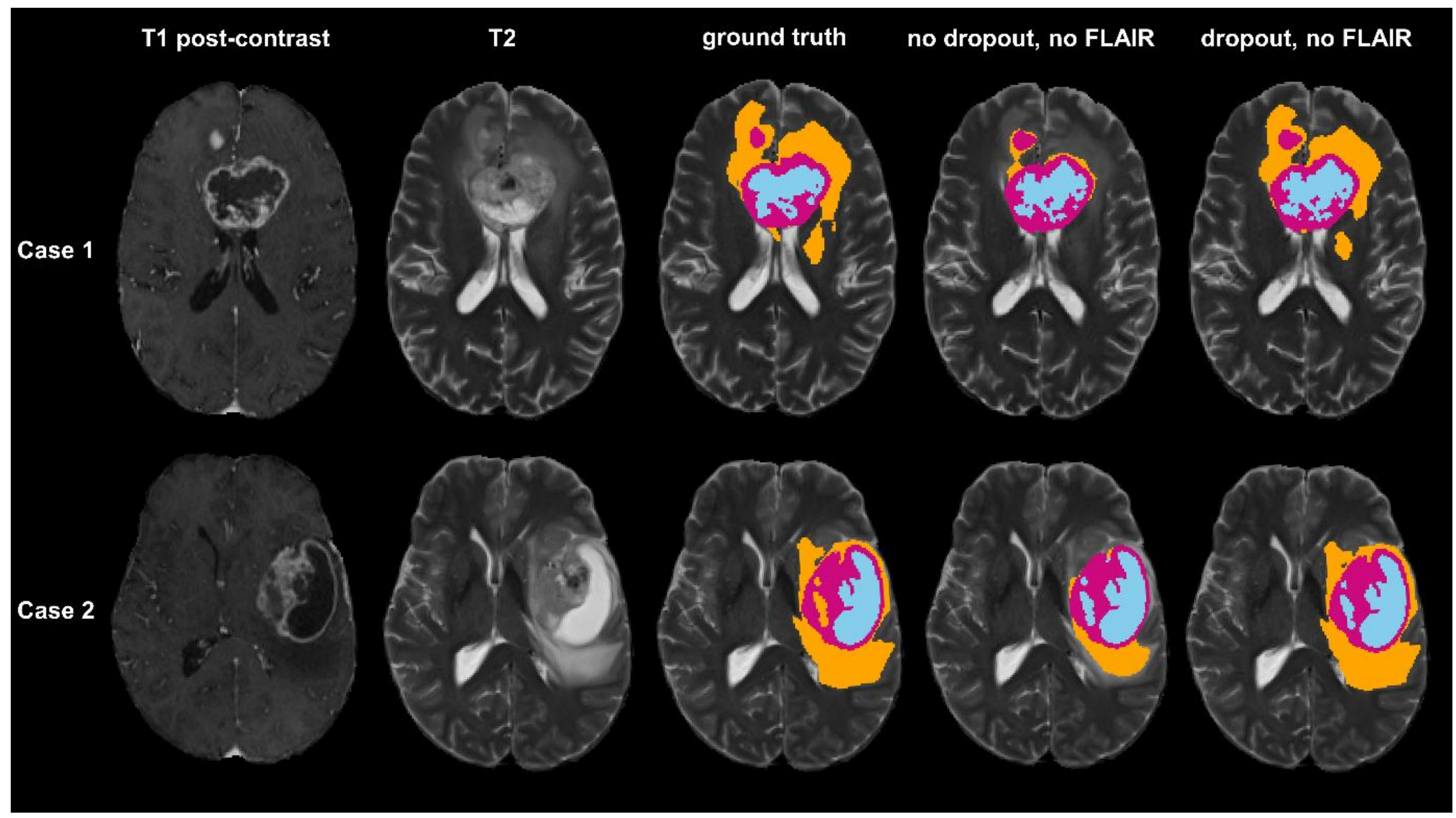

**Figure 4. Representative external validation cases showing recurrent whole-tumor and edema rescue without T2-FLAIR.** For each case, post-contrast T1-weighted and T2-FLAIR images are shown together with the reference standard, the region-wise nnU-Net prediction without dropout (r=0.0) in the T2-FLAIR-absent inference scenario, and the prediction with targeted T2-FLAIR dropout (r=0.35) in the same scenario. In the T2-FLAIR-absent columns, the T2-FLAIR channel was replaced with zeros at inference. These cases illustrate recurrent edema and whole-tumor undersegmentation without dropout and recovery with targeted dropout, corresponding to the cohort-level whole-tumor volumetric bias shown in Figure 2.

**Abbreviations:** FLAIR = fluid-attenuated inversion recovery; r = dropout rate.

## Supplementary Tables

| | Overall | | Enhancing Tumor | | Non-enhancing Tumor | | Edema | |
|---|---|---|---|---|---|---|---|---|
| T2-FLAIR drop for training | DSC med [Q1-Q3] | HD95 med [Q1-Q3] | DSC med [Q1-Q3] | HD95 med [Q1-Q3] | DSC med [Q1-Q3] | HD95 med [Q1-Q3] | DSC med [Q1-Q3] | HD95 med [Q1-Q3] |
| *Segmentation with T2-FLAIR input* | | | | | | | | |
| r=0.0 | 91.0 [82.0-95.0] | 2.01 [1.07-3.46] | 94.0 [88.0-97.0] | 1.00 [1.00-1.41] | 91.0 [79.0-97.0] | 1.41 [1.00-4.12] | 92.0 [85.0-96.0] | 1.73 [1.00-3.46] |
| r=0.35 | 91.0 [81.0-95.0] | 2.00 [1.14-3.42] | 94.0 [87.0-97.0] | 1.00 [1.00-1.41] | 92.0 [79.0-97.0] | 1.41 [1.00-4.12] | 92.0 [85.0-96.0] | 1.41 [1.00-3.61] |
| r=0.5 | 90.0 [81.0-95.0] | 2.04 [1.14-3.58] | 94.0 [87.0-97.0] | 1.00 [1.00-1.41] | 91.0 [79.0-97.0] | 1.41 [1.00-4.24] | 91.0 [85.0-95.0] | 1.72 [1.00-3.61] |
| *Segmentation without T2-FLAIR input* | | | | | | | | |
| r=0.0 | 64.0 [55.0-71.0] | 9.17 [6.53-13.06] | 91.0 [84.0-95.0] | 1.41 [1.00-2.24] | 89.0 [72.0-96.0] | 2.24 [1.00-5.20] | 14.0 [3.0-28.0] | 22.44 [16.43-31.03] |
| r=0.35 | 89.0 [79.0-93.0] | 2.26 [1.41-4.06] | 94.0 [87.0-96.0] | 1.00 [1.00-1.41] | 92.0 [79.0-97.0] | 1.73 [1.00-4.24] | 87.0 [78.0-92.0] | 2.45 [1.73-4.69] |
| r=0.5 | 89.0 [79.0-94.0] | 2.30 [1.38-4.17] | 93.0 [87.0-96.0] | 1.00 [1.00-1.41] | 92.0 [76.0-97.0] | 1.73 [1.00-4.36] | 88.0 [79.0-92.0] | 2.24 [1.73-4.64] |

**Supplementary Table S1. UPenn-GBM external validation (n=403): Label-wise segmentation performance with T2-FLAIR present and absent at inference.** Performance of label-wise nnU-Net models (NCR/NET, ED, ET) trained on BraTS 2021 with targeted T2-FLAIR dropout probability rates of 0.0, 0.35, and 0.50 and evaluated under two prespecified inference scenarios: T2-FLAIR present and T2-FLAIR absent (T2-FLAIR channel replaced with zeros). Values are median [Q1-Q3] across patients. Overall denotes the unweighted macro-average across NCR/NET, ED, and ET.

**Abbreviations:** DSC = Dice similarity coefficient (reported as %); HD95 = 95th-percentile Hausdorff distance (mm); NCR/NET = necrotic and non-enhancing tumor; ED = edema; ET = enhancing tumor; FLAIR = fluid-attenuated inversion recovery; r = dropout rate.

| | Median DSC r(0.35)-r(0) [95%CI] | TOST p-Value |
|---|---|---|
| *Segmentation with T2-FLAIR input* | | |
| Overall | 0.0442 [0.002, 0.092] | <0.001 |
| ET | -0.0071 [-0.056, 0.027] | <0.001 |
| NET | 0.0320 [0.011, 0.074] | <0.001 |
| Edema | -0.0417 [-0.078, 0.029] | <0.001 |
| *Segmentation without T2-FLAIR input* | | |
| Overall | 24.1974 [23.613, 24.685] | 1 |
| ET | 1.075 [0.916, 1.241] | 1 |
| NET | 0.9508 [0.705, 1.213] | 1 |
| Edema | 69.3675 [67.202, 70.926] | 1 |

**Supplementary Table S2. UPenn-GBM external validation (n=403): Paired equivalence testing (TOST) for r=0.35 versus r=0.0 in label-wise models.** Label-wise models trained with targeted T2-FLAIR dropout (r=0.35) were compared with no-dropout models (r=0.0) under T2-FLAIR-present and T2-FLAIR-absent inference scenarios. Entries report median paired DSC differences (r=0.35 − r=0.0) in percentage points with 95% bootstrap confidence intervals. TOST p values test equivalence within prespecified bounds of ±1.5 DSC percentage points (α=0.05).

**Abbreviations:** TOST = two one-sided tests; DSC = Dice similarity coefficient; NCR/NET = necrotic and non-enhancing tumor; ED = edema; ET = enhancing tumor; FLAIR = fluid-attenuated inversion recovery; r = dropout rate.

| | Median DSC r(0.35)-r(0.50) [95%CI] | TOST p-Value |
|---|---|---|
| *Segmentation with T2-FLAIR input* | | |
| Overall | 0.075 [0.049, 0.093] | <0.001 |
| ET | 0.034 [0.003, 0.066] | <0.001 |
| TC | 0.049 [0.028, 0.067] | <0.001 |
| WT | 0.075 [0.038, 0.108] | <0.001 |
| *Segmentation without T2-FLAIR input* | | |
| Overall | -0.113 [-0.146, -0.076] | <0.001 |
| ET | -0.061 [-0.10, -0.006] | <0.001 |
| TC | -0.035 [-0.054, -0.018] | <0.001 |
| WT | -0.201 [-0.233, -0.140] | <0.001 |

**Supplementary Table S3. UPenn-GBM external validation (n=403): Paired equivalence testing (TOST) for r=0.35 versus r=0.50 in region-wise models.** Region-wise models trained with targeted T2-FLAIR dropout (r=0.35 and r=0.50) were compared under T2-FLAIR-present and T2-FLAIR-absent inference scenarios. Entries report median paired DSC differences (r=0.35 − r=0.50) in percentage points with 95% bootstrap confidence intervals. TOST p values test equivalence within prespecified bounds of ±1.5 DSC percentage points (α=0.05).

**Abbreviations:** TOST = two one-sided tests; DSC = Dice similarity coefficient; WT = whole tumor; TC = tumor core; ET = enhancing tumor; FLAIR = fluid-attenuated inversion recovery; r = dropout rate.

| | Median DSC r(0.35)-r(0.50) [95%CI] | TOST p-Value |
|---|---|---|
| *Segmentation with T2-FLAIR input* | | |
| Overall | 0.247 [0.183, 0.316] | <0.001 |
| ET | 0.188 [0.118, 0.233] | <0.001 |
| NET | 0.256 [0.171, 0.329] | <0.001 |
| Edema | 0.172 [0.122, 0.218] | <0.001 |
| *Segmentation without T2-FLAIR input* | | |
| Overall | -0.051 [-0.138, 0.013] | <0.001 |
| ET | -0.007 [-0.067, 0.041] | <0.001 |
| NET | 0.0484 [0.000, 0.136] | <0.001 |
| Edema | -0.234 [-0.301, -0.172] | <0.001 |

**Supplementary Table S4. UPenn-GBM external validation (n=403): Paired equivalence testing (TOST) for r=0.35 versus r=0.50 in label-wise models.** Label-wise models trained with targeted T2-FLAIR dropout (r=0.35 and r=0.50) were compared under T2-FLAIR-present and T2-FLAIR-absent inference scenarios. Entries report median paired DSC differences (r=0.35 − r=0.50) in percentage points with 95% bootstrap confidence intervals. TOST p values test equivalence within prespecified bounds of ±1.5 DSC percentage points (α=0.05).

**Abbreviations:** TOST = two one-sided tests; DSC = Dice similarity coefficient; NCR/NET = necrotic and non-enhancing tumor; ED = edema; ET = enhancing tumor; FLAIR = fluid-attenuated inversion recovery; r = dropout rate.

| BraTS Label/Region | HD-Glio Label | r=0.35 Median DSC (%) [Q1-Q3] | HD-GLIO Median DSC (%) [Q1-Q3] | Median Δ DSC (%) (model - HD) | Non-inferior p-Value |
|---|---|---|---|---|---|
| Edema | NE | 91.77 [84.91-95.55] | 87.68 [78.30-92.30] | 3.158 [2.733, 3.792] | <0.001 |
| ET | CE | 93.96 [87.49-96.80] | 90.31 [83.79-93.40] | 3.026 [2.698, 3.262] | <0.001 |
| WT | NE+CE | 95.65 [91.94-97.81] | 91.01 [84.77-94.24] | 4.087 [3.702, 4.360] | <0.001 |

**Supplementary Table S5. UPenn-GBM external validation (n=403): Noninferiority of the r=0.35 dropout model versus HD-GLIO in the T2-FLAIR-present scenario.** The targeted T2-FLAIR dropout model (r=0.35) was compared with HD-GLIO under the T2-FLAIR-present inference scenario. Endpoints were compared using the following prespecified correspondences: ET (BraTS) versus CE (HD-GLIO), ED (BraTS label-wise edema) versus NE (HD-GLIO), and WT (BraTS region-wise) versus NE+CE (HD-GLIO). Values are median DSC (%) [Q1-Q3]. Median ΔDSC denotes the paired median difference (model − HD-GLIO) in DSC percentage points. Noninferiority p values correspond to a one-sided test with a prespecified noninferiority margin of −1.5 DSC percentage points (α=0.05).

**Abbreviations:** DSC = Dice similarity coefficient; ET = enhancing tumor; ED = edema; WT = whole tumor; CE = contrast enhancing; NE = non-enhancing; r = dropout rate.

## Supplementary File: CLAIM Checklist (2024 Update)

Manuscript: **Robust Glioblastoma Segmentation and Volumetry Without T2-FLAIR: External Validation of Targeted Dropout Training**

| CLAIM item | Checklist requirement (paraphrased) | Addressed | Manuscript location | Notes |
|---|---|---|---|---|
| 1 | Identify the work as an AI/ML study in the title/abstract. | Yes | Title; Abstract header and text | |
| 2 | Provide a structured abstract that summarizes design, data, model, evaluation, and main results. | Yes | Abstract (Purpose/Materials and Methods/Results/Conclusion) | |
| 3 | Describe clinical/scientific background and intended use/role of the AI tool. | Yes | Introduction | |
| 4 | State study objective(s) and/or hypothesis(es). | Yes | End of Introduction (Purpose and hypotheses) | |
| 5 | Specify whether the study is prospective or retrospective. | Yes | Methods – Study design and reporting | |
| 6 | State study goal (e.g., model development, validation, noninferiority/equivalence, etc.). | Yes | Introduction, Methods | |
| 7 | Describe data sources (origin, dataset name(s), public/private, recruitment). | Yes | Methods – Datasets and reference standard (BraTS 2021; UPenn-GBM) | |
| 8 | Define inclusion/exclusion criteria and study setting/time frame. | Yes | Methods – Datasets and reference standard | |
| 9 | Describe preprocessing steps applied to the data. | Yes | Methods – Datasets and reference standard | BraTS organizer preprocessing is described |
| 10 | Describe any selection of subsets/regions/patches/frames for model input. | Yes | Methods – Training and testing | Extraction of UPenn GBM dataset for external validation dataset is described |
| 11 | Describe de-identification and privacy protection approach. | Yes | Methods – Study design and reporting | Only public de-identified datasets are used |
| 12 | Describe missing data and how it was handled. | NA | Methods | Full public dataset was used, which only contains subjects with full four modality imaging set |
| 13 | Provide imaging acquisition protocol details (scanner, sequences, key parameters) or cite where available. | Partial | Methods | BraTS is a heterogenous imaging dataset, parameters for each site are publicly available and referenced |
| 14 | Identify and justify the reference standard used. | Yes | Methods | |
| 15 | Describe how reference annotations/labels were generated. | Yes | Methods | |

| CLAIM item | Checklist requirement (paraphrased) | Addressed | Manuscript location | Notes |
|---|---|---|---|---|
| 16 | Define what constitutes a positive case / target definition(s). | Yes | Methods | Tumor subcompartments and derived regions are defined |
| 17 | Describe reference standard quality assurance (expertise, blinding, adjudication). | Yes | Methods | |
| 18 | Report inter- and/or intra-rater variability (and/or mitigation) for the reference standard. | NA | | Ground truth segmentations obtained from BraTS dataset |
| 19 | Describe data partitioning for training/validation/testing (and sizes). | Yes | Methods | |
| 20 | State the level of disjointness between partitions (e.g., patient-level). | NA | | |
| 21 | Describe how sample size was determined (power/precision or rationale). | Partial | Methods | No formal sample size calculation was performed, as all available data were used |
| 22 | Describe model architecture; cite prior architecture if used; describe modifications. | Yes | Methods | nnU-Net 3D full-resolution used, with region- and label-based training, with and without FLAIR-dropout at fixed proportions |
| 23 | Report software/frameworks and versions; provide access information when applicable. | Yes | Methods | |
| 24 | Describe model parameter initialization (including randomness/seed control). | | | |
| 25 | Describe training procedure and hyperparameters (optimizer, LR, epochs, augmentation, loss, early stopping). | Partial | Methods | Epochs and use of nnU-Net defaults are stated; consider adding a concise hyperparameter summary or nnU-Net plan/config file in Supplementary Methods. |
| 26 | Describe how the final model/checkpoint was selected. | Yes | Methods | |
| 27 | Describe any ensembling and how outputs were combined. | Yes | Methods | |
| 28 | Specify evaluation metrics and rationale/clinical relevance. | Yes | Methods – Statistics | |
| 29 | Describe statistical analysis and uncertainty estimation. | Yes | Methods – Statistics | |
| 30 | Describe robustness/sensitivity analyses. | Yes | Methods - Statistics | |
| 31 | Describe model explainability/interpretability methods (if used). | NA | | |

| CLAIM item | Checklist requirement (paraphrased) | Addressed | Manuscript location | Notes |
|---|---|---|---|---|
| 32 | Report performance on internal data (training/validation/test as applicable). | NA | | |
| 33 | Report performance on external data (if applicable). | Yes | Results; Supplementary | |
| 34 | Provide clinical trial registration information (if applicable). | NA | | Retrospective analysis of public de-identified datasets; no registration expected. |
| 35 | Provide participant/data flow (numbers included/excluded) per cohort. | Yes | Methods | |
| 36 | Report participant demographics/clinical characteristics (when available). | Partial | Methods | BraTS demographics are briefly described and cited |
| 37 | Report performance metrics with measures of uncertainty (CI, IQR) and appropriate aggregation. | Yes | Results and Supplementary | Medians/IQRs and bootstrap CIs for paired differences are provided. |
| 38 | Report subgroup analyses (if performed) and pre-specification. | NA | | |
| 39 | Provide analysis of errors/failures (qualitative or quantitative). | Yes | Results/Discussion | |
| 40 | Report interpretability outputs/results (if interpretability methods used). | NA | | |
| 41 | Discuss limitations, biases, and generalizability. | Yes | Discussion – Limitations paragraph | |
| 42 | Discuss clinical implications and future work. | Yes | Discussion; Conclusion | |
| 43 | State availability of data, code, and/or trained model (or restrictions). | No | Data availability statement | |
| 44 | Report funding sources and role of funders. | Yes | Funding statement | |